\documentclass[aps,pre,reprint,groupedaddress,showpacs]{revtex4-1}

\usepackage[english]{babel}
\usepackage[latin2]{inputenc}

\usepackage{amsmath}
\usepackage{amssymb}
\usepackage{graphicx}
\usepackage{verbatim}

\begin{document}

\title{Thermodynamically consistent Langevin dynamics with spatially correlated noise predicts frictionless regime and transient attraction effect.}

\author{M. Majka}
\email{maciej.majka@uj.edu.pl}
\affiliation{Marian Smoluchowski Institute of Physics, Jagiellonian University, ul. prof. Stanis\l{}awa \L{}ojasiewicza 11, 30-348 Krak\'{o}w Poland}
\author{P. F. G\'{o}ra}
\affiliation{Marian Smoluchowski Institute of Physics, Jagiellonian University, ul. prof. Stanis\l{}awa \L{}ojasiewicza 11, 30-348 Krak\'{o}w Poland}

\begin{abstract}
While the origin of temporal correlations in Langevin dynamics have been thoroughly researched, the understanding of Spatially Correlated Noise (SCN) is rather incomplete. In particular, very little is known about the relation between friction and SCN. In this article, we derive the formal formula for the spatial correlation function in the particle-bath interactions. This expression shows that SCN is the inherent component of binary mixtures, originating from the effective (entropic) interactions. Further, employing this spatial correlation function, we postulate the thermodynamically consistent Langevin equation driven by SCN and the adequate Fluctuation-Dissipation Relation. The thermodynamical consistency is achieved by introducing the spatially variant friction coefficient, which can be also derived analytically. This coefficient exhibits a number of intriguing properties, e.g. the singular behavior for certain interaction types. Eventually, we apply this new theory to the system of two charged particles in the presence of counter-ions. Such particles interact via the screened-charge Yukawa potential and the inclusion of SCN leads to the emergence of the anomalous frictionless regime. In this regime the particles can experience active propulsion leading to the transient attraction effect. This effect suggests a non-equilibrium mechanism facilitating the molecular binding of the like-charged particles. 
\end{abstract}

%05.40.-a - Fluctuation phenomena, random processes, noise, and Brownian motion
%05.40.Ca - noise 
%82.70.Dd - colloids
%87.16.dr - Assembly and interactions
%87.15.Vv - diffusion

\maketitle

\section{Introduction}
Spatially Correlated Noise (SCN) is encountered in e.g. the transport in plasma \cite{bib:plasma1,bib:plasma2}, active particle motion \cite{bib:active}, phase transitions \cite{bib:phasetrans}, glass dynamics \cite{bib:glass1,bib:glass2,bib:glass3}, cytoplasmic micro-flows \cite{bib:cyto1,bib:cyto2} etc. The review of the SCN-related phenomena can be found in Ref. \cite{bib:majka0}. While the time-correlated noise and the related Generalized Langevin Equations are well understood on the microscopic basis thanks to the Mori-Zwanzig theory \cite{bib:zwanzig}, little has been done to formalize the dynamics of SCN-driven systems. No general microscopic model for SCN has been proposed, no thermodynamically consistent Langevin equation with SCN has been formulated and the Fluctuation-Dissipation Relation (FDR) for SCN (i.e. the interdependence of the correlations in noise and friction) is neither known. While many papers report interesting effects related to SCN (e.g \cite{bib:majka1,bib:majka2,bib:denisov,bib:gora,bib:neurons,bib:tkachenko}), they all relay on the generic Stokesian friction. Certain insight into the origin of SCN can be obtained via the two-particle Mori-Zwanzig model \cite{bib:majka0}, but it is limited to short distances and the linear particle-bath coupling.

In this article we attempt to establish the thermodynamically consistent theory of SCN. First, in Section \ref{sec:corr} we show that SCN is inherently present in the two-component (binary) mixtures due to the effective interactions \cite{bib:likos,bib:lekkerkerker,bib:majka3}. This ubiquity of SCN, especially in the typical soft-matter systems, has not been recognized before. In the framework of binary mixture theory we establish the exact relation between the correlation of particle-bath interaction and the correlation of effective forces. Knowing this relation, in Section \ref{sec:Langevin} we formulate the Langevin dynamics which both utilizes SCN and leads to the Boltzmann distribution in the steady-state. These two constraints require that the Spatially Variant Friction Coefficient (SVFC) is postulated. We derive SVFC analytically for arbitrary potentials and thus we establish FDR for SCN. In Section \ref{sec:svfc_gen_prop} we discuss the general properties of SVFC such as the emergence of the frictionless regime for short distances, a possible singular behavior and the attraction arising in its vicinity.

Finally, in Section \ref{sec:DLVO}, we apply this new dynamics to the system of two charged spheres, with repulsion screened by counter-ions. According to the Derjaguin-Landau-Vervey-Overbeek (DLVO) theory such particles repel via the Yukawa potential \cite{bib:lekkerkerker,bib:majka3,bib:crocker}. This is a popular model in soft matter \cite{bib:lekkerkerker}, bio- \cite{bib:prot1,bib:prot2} and plasma \cite{bib:plasma3} physics. Indeed, the inclusion of SCN leads to the concentration-controlled emergence of the frictionless regime at whose border the bath-induced propulsion occurs. This results in the transient attraction effect, i.e. for certain initial conditions the particles can actively overcome the electrostatic repulsion. The attraction of the like-charged spheres is a well-known phenomenon \cite{bib:like1, bib:like2, bib:like3}, but with no consensus on its exact explanation \cite{bib:like_expl1, bib:like_expl2, bib:like_expl3}. Our model suggests that the non-equilibrium contribution might be also important for this effect and SCN can account for another mechanism that facilitates the self-assembly and molecular transport. In Section \ref{sec:caveats} we provide the critical summary of these results.

%--------------------------CORRELATION FUNCTION DERIVATION--------------------
\section{Spatial correlations in the tracer-bath interactions}\label{sec:corr}
First, we will show that the particle-bath interaction in the binary mixture is spatially correlated. Let us consider $N$ 'tracer' particles at positions $x_i$ in the thermal bath of $\tilde N$ depletant particles at positions $\tilde x_i$.  The respective fugacities are $z$ and $\tilde z$. The system is one dimensional and it has a finite size $\Omega$, which is much greater than the range of interactions. There are three microscopic potentials in the system: tracer-tracer interaction $U_0(x_i-x_j)$, tracer-bath interaction $V(x_i-\tilde x_j)$ and (optionally) bath-bath interaction $v(\tilde x_i-\tilde x_j)$. The partition function of this system reads:
\begin{equation}
\Xi=\frac{z^N}{N!}\int_{\Omega} \{dx\}\exp\left(-\beta\sum_{i>j}^N U_0(x_i-x_j)\right) \tilde \Xi
\end{equation}
where $\{d x\}=\prod_n^{ N} d{x}_n$ and $\tilde \Xi$ reads:
\begin{equation}
\tilde \Xi=\frac{{\tilde z}^{\tilde N}}{\tilde N!}\int_{\Omega} \{d\tilde x\}\exp\left(-\beta \sum_i^{N}\sum_j^{\tilde N}V(x_i-\tilde x_j)-\beta \tilde v\right) \label{eq:tildeXi}
\end{equation}
with $\tilde v=\sum_{k>l}^{\tilde N}v(\tilde x_k-\tilde x_l)$ used for brevity. $\tilde \Xi$ is the source of effective interactions, which we restrict to the two-body terms \cite{bib:likos,bib:majka3}:
\begin{equation}
-\frac{1}{\beta}\ln \tilde \Xi \simeq \sum_{i>j}^N U_{eff}(x_i-x_j) \label{eq:Ueff_def}
\end{equation}
 The microscopic force exerted on the $i$-th tracer by the bath reads:
\begin{equation}
\xi(x_i)=-\sum_{j}^{\tilde N}\partial_{x_i}V(x_i-\tilde x_j) \label{eq:coupling}
\end{equation}
We want to know its covariance $<\xi(0)\xi(r)>$, which formally reads:
\begin{equation}
\begin{split}
&<\xi(0)\xi(r)>=\frac{z^N{\tilde z}^{\tilde N}}{N!{\tilde N}!\Xi}\int_\Omega \{dx\}\{d\tilde x\} \xi(x_i)\xi(x_i+r)\times\\
&\times e^{-\beta\left( \sum_{k>l}^N U_0(x_k-x_l)+\sum_k^N\sum_l^{\tilde N}V(x_k-\tilde x_l)+\tilde v\right)}
\end{split}\label{eq:cov_formal}
\end{equation}
First, we will derive a useful identity. Let us start with the following observation:
\begin{equation}
\sum_j^{\tilde N}V(x_i-\tilde x_j) =\int_\Omega dy \rho_i(y)\xi(y)
\end{equation}
where $\rho_i(y)=\theta(y-x_i)$ is the Heavyside step function, satisfying $\partial_y\rho_i(y)=\delta(y-r_i)$. We can now treat $\rho_i(y)$ as a density field (in fact: the integrated single-particle density) and write:
\begin{equation}
\begin{split}
&\xi(x_i)\xi(x_i+r)\exp\left(-\beta\sum_j^{\tilde N}V(x_i-\tilde x_j) \right)=\\
&=\frac{1}{\beta^2}\frac{\delta^2}{\delta\rho_i(x_i)\delta \rho_i(x_i+r)}\exp\left(-\beta\int_{\Omega} dx\rho_i(y)\xi(y) \right)\\
&=\frac{1}{\beta^2}\frac{\delta^2}{\delta\rho_i(x_i)\delta \rho_i(x_i+r)}\exp\left(-\beta\sum_j^{\tilde N}V(x_i-\tilde x_j) \right)
\end{split}\label{eq:id}
\end{equation}
where $\frac{\delta}{\delta \rho_i(x_i)}$ denotes the functional derivative \cite{bib:hansen, bib:phasetrans}. Applying \eqref{eq:id} in \eqref{eq:cov_formal}, we obtain:
\begin{equation}
\begin{split}
&<\xi(0)\xi(r)>=\frac{z^N {\tilde z}^{\tilde N}}{N!{\tilde N}! \Xi}\int_\Omega \{dx\}\{d\tilde x\} e^{-\beta\sum_{k>l}^N U_0(x_k-x_l)}\times \\
&\times \frac{1}{\beta^2}\frac{\delta^2}{\delta\rho_i(x_i)\delta \rho_i(x_i+r)}e^{-\beta\left(\sum_k^N\sum_l^{\tilde N}V(x_k-\tilde x_l)+\tilde v\right)}
\end{split}
\end{equation}
Exchanging the order of the integrations $\int_\Omega \{d\tilde{r}\}$ and the functional derivatives, by the definitions \eqref{eq:tildeXi} and \eqref{eq:Ueff_def}, we arrive at:
\begin{equation}
\begin{split}
&<\xi(0)\xi(r)>=\frac{z^N}{N!\beta^2 \Xi}\int_\Omega \{dx\}e^{-\beta\sum_{k>l}^N U_0(x_k-x_l)}\times \\
&\times\frac{\delta^2}{\delta\rho_i(x_i)\delta \rho_i(x_i+r)}e^{-\beta\sum_{k>l}^N U_{eff}(x_k-x_l)} \label{eq:intermediate1}
\end{split}
\end{equation}
Although the calculation of the effective potential can be complicated, we know where the distribution $\rho_i(y)$ must reappear \cite{bib:majka3}, namely:
\begin{equation}
\sum_{k\neq i}^NU_{eff}(x_i-x_k)=\int_\Omega dx (\partial_y\rho_i(y))\sum_{k\neq i}^NU_{eff}(y-x_k)
\end{equation}
This allows us to perform the functional derivatives in \eqref{eq:intermediate1}, to eventually obtain:
\begin{equation}
\begin{split}
&<\xi(0)\xi(r)>=\frac{z^N}{N!\Xi}\int_\Omega \{dx\}e^{-\beta \sum_{k>l}^N U(x_k-x_l)}\times \\
&\times \left(\sum_{j\neq i}^N F_{eff}(x_i-x_j)\right)\left(\sum_{l\neq i}^N F_{eff}(x_i-x_l+r)\right) \label{eq:spat_cov}
\end{split}
\end{equation}
where:
\begin{gather}
U(x_k-x_l)=U_0(x_k-x_l)+U_{eff}(x_k-x_l) \\ 
F_{eff}(x_i-x_k)=-\partial_{x_i}U_{eff}(x_i-r_k)
\end{gather}
In our considerations we assume that $\Omega$ is finite, because for realistic potentials such that $U(r\to+\infty)=0$, $\Xi$ is divergent as $\Omega\to+\infty$. This makes the covariance ill-defined in the thermodynamic limit. However, the correlation function:
\begin{equation}
h(r)=<\xi(0)\xi(r)>/<\xi^2(0)>  \label{eq:spat_corr}
\end{equation}
is not affected by this problem, because the divergent $\Xi$ in \eqref{eq:spat_corr} cancels out. Therefore, $h(r)$ always exists, provided that the integral in \eqref{eq:spat_cov} is finite.

The result \eqref{eq:spat_cov} shows that the tracer-bath interaction $\xi(r)$ is spatially correlated and this correlation depends directly on the effective forces $F_{eff}(r)$. Our considerations are valid for any type of binary mixture, suggesting that the spatial correlations in $\xi(r)$ are a ubiquitous phenomenon.

%-------------LANGEVIN EQUATIONS DERIVATION------------------------------------
\section{Thermodynamically consistent Langevin equations with SCN}\label{sec:Langevin}
Having found $h(r)$, we will now formulate the SCN-driven Langevin dynamics. The Langevin equations provide an insight into both the equilibrium and non-equilibrium regime. However, the agreement between their predictions and the microscopic physics (given by e.g. Boltzmann distribution) can be assured only for the equilibrium state. In fact, there might exist many different types of Langevin dynamics that lead to the same equilibrium distribution, but differ e.g. in the properties of the stochastic force and their transient behavior \cite{bib:majka0}. However, one might assume that the more microscopic properties Langevin dynamics reproduces, the closer it is to the actual non-equilibrium physics. 

The ordinary Langevin dynamics replaces the deterministic tracer-bath coupling \eqref{eq:coupling} with the stochastic force and friction, but assumes that this force is completely uncorrelated. This dynamics is well-known to reproduce the Boltzmann distribution of tracers in the steady-state \cite{bib:gardiner}. However, it does not specify whether the effective interactions are included or not. On the other hand, the binary mixture perspective clearly indicates that the Boltzmann distribution must contain the effective interactions. These interactions also mean that the force of tracer-bath coupling $\xi(x_i)$ is spatially correlated, according to $h(r)$. 

The SCN-driven Langevin dynamics  must reproduce this fact, i.e. we demand that the stochastic force (replacing the microscopic particle-bath coupling) is also spatially correlated according to $h(r)$. Since $h(r)$ has been derived from the Boltzmann distribution, we can expect that this distribution is recovered in the steady-state limit. Our calculations are carried out in the over-damped regime. From now on, $\xi_i$ denotes the correlated Gaussian noise affecting the $i$-th tracer, which satisfies:
\begin{gather}
\begin{aligned}
&<\xi_i(t)>=0&&<\xi_i^2(t)>=\sigma^2 
\end{aligned} \\
<\xi_i(t)\xi_j(t')>=\sigma^2 \delta(t-t') h(x_i-x_j)
\end{gather}
Further, we shall omit the time $t$ as an argument of $\xi_i$, since it is of no interest here. We consider a system of two tracers which is one-dimensional and has a huge, though finite volume $\Omega$. Let us start with the ordinary over-damped equations of motion, driven by SCN:
\begin{equation}
\gamma \dot x_i=F(x_i-x_j)+\xi_i  \label{eq:eq_init}
\end{equation}
where $\gamma$ is a constant friction coefficient and $F(r)=F_0(r)+F_{eff}(r)$. Switching to the relative distance $r=x_2-x_1$ and the position of the mass center $R=(x_1+x_2)/2$, the equations of motion read:
\begin{gather}
\gamma \dot r=2F(r)+\xi_2-\xi_1 \\
\gamma \dot R=\frac{1}{2}(\xi_2+\xi_1)
\end{gather}
It is known that the linear combination of the Gaussian variables can be replaced with another properly rescaled Gaussian variable. Namely:
\begin{equation}
\xi_2\pm\xi_1=\sqrt{2}\sigma\sqrt{1\pm h(r)}\eta_{\pm} \label{eq:eta_def}
\end{equation}
where $\eta_{\pm}$ is now drawn from the normal distribution, $<\eta_+\eta_->=0$ and $<\eta_{\pm}(t)\eta_{\pm}(t')>=\delta(t-t')$. One might check that:
\begin{equation}
<(\xi_2\pm\xi_1)^2>=2\sigma^2 (1\pm h(r))<\eta_{\pm}^2>
\end{equation}
For brevity we will denote:
\begin{equation}
g_{\pm}(r)=\sqrt{1\pm h(r)}
\end{equation}
Under the substitution \eqref{eq:eta_def}, the equations of motion turn into:
\begin{gather}
\gamma \dot r=2F(r)+\sqrt{2}\sigma g_{-}(r)\eta_{-} \\
\gamma \dot R=\frac{\sigma}{\sqrt{2}} g_{+}(r)\eta_{+}
\end{gather}
thus for $h(r)\neq0$ the noise becomes multiplicative. Let us specify that we will use the Stratonovich interpretation \cite{bib:gardiner} everywhere, which is typical for diffusion. We can now write down the Fokker-Planck Equation (FPE) for this system, which, in the stationary limit, reads:
\begin{equation}
\begin{split}
&-\partial_r\left(\frac{2F(r)}{\gamma}\tilde P_{st} \right)+\sigma^2\partial_{r}\left[\frac{g_{-}(r)}{\gamma}\partial_r \left(\frac{g_{-}(r)}{\gamma}  \tilde P_{st}\right)\right]+\\
&+\frac{\sigma^2}{4} g_{+}^2(r)\partial_{RR}\tilde P_{st}=0
\end{split}\label{eq:FPE_init}
\end{equation}
For the non-correlated case $h(r)=0$ and $g_{\pm}(r)=1$, in which case $\gamma=\beta\sigma^2/2$, the solution of this equation reads:
\begin{equation}
P_{B}(r,R)=\mathcal{N}^{-1} \exp\left(\beta\int^r_0 dr' F(r')\right) \label{eq:P_Boltz}
\end{equation}
This is the Boltzmann distribution with normalization constant $\mathcal{N}$. For clarity, by $P_B(r,R)$ we mean that the equilibrium probability distribution of $R$ is simply homogeneous in $\Omega$. 

For $h(r)\neq 0$, $P_B(r,R)$ is no longer a valid solution to \eqref{eq:FPE_init}, in which case $\tilde P_{st}(r,R)$ reads:
\begin{equation}
\tilde P_{st}(r,R)=\exp\left(\frac{2\gamma}{\sigma^2}\int_0^r dr' \frac{F(r')}{g_{-}^2(r')}-\ln \frac{g_{-}(r)}{\gamma}+\tilde C \right)
\end{equation}
One can see that for $g_{-}(r)\neq1$ it is impossible that $P_B(r,R)=\tilde P_{st}(r,R)$, i.e. the presence of SCN leads to the non-Boltzmann equilibrium distribution. However, since $h(r)$ agrees with equilibrium state, there is no reason to expect that Boltzmann distribution is violated. This means that our theory is not thermodynamically consistent. The simplest way to resolve this issue is to introduce the Spatially Variant Friction Coefficient (SVFC) $K(r)$, namely we postulate that the equations of motion read:
\begin{gather}
K(r) \dot r=2F(r)+\sqrt{2}\sigma g_{-}(r)\eta_{-} \label{eq:r2} \\
\frac{\gamma}{2} \dot R=\frac{\sigma}{\sqrt{2}} g_{+}(r)\eta_{+} \label{eq:R2}
\end{gather}
In other words, we assume that the motion of one particle affects the friction coefficient of the other particle and vice versa. The stationary FPE equation for this system reads:
\begin{equation}
\begin{split}
&-\partial_r\left(\frac{2F(r)}{K(r)}P_{st} \right)+\sigma^2\partial_{r}\left[\frac{g_{-}(r)}{K(r)}\partial_r \left(\frac{g_{-}(r)}{K(r)}  P_{st} \right)\right]+\\
&+\frac{\sigma^2}{4} g_{+}^2(r)\partial_{RR}P_{st}=0
\end{split} \label{eq:eq_Pst}
\end{equation}
Assuming no probability current, the exact solution of \eqref{eq:eq_Pst} is given by:
\begin{equation}
P_{st}(r,R)=\exp\left(\frac{2}{\sigma^2}\int_0^r dr' \frac{F(r')K(r')}{g_{-}^2(r')}-\ln \frac{g_{-}(r)}{K(r)}+C \right) \label{eq:Pstat}
\end{equation}
We now explicitly demand that:
\begin{equation}
P_B(r,R)=P_{st}(r,R) \label{eq:PB=Pst}
\end{equation}
Differentiating both sides of \eqref{eq:PB=Pst} with respect to $r$ and rearranging we arrive at the equation for $K(r)$:
\begin{equation}
K'(r)-\left(\beta F(r)+\frac{{g'}_{-}(r)}{g_{-}(r)} \right)K(r)=-\frac{2}{\sigma^2}\frac{F(r)}{g_{-}^2(r)}K^2(r) \label{eq:eq_K}
\end{equation}
This is the Bernoulli differential equation and it can be solved exactly. Applying the boundary condition $K(r\to+\infty)=\gamma$, which means that far away both tracers experience the Stokes-like friction, we obtain:
\begin{equation}
K(r)=\frac{g_{-}(r)e^{-\beta U(r)}}{\frac{1}{\gamma}-\frac{2}{\sigma^2}\int_r^{+\infty}dr'\frac{F(r')}{g_{-}(r')}e^{-\beta U(r')}} \label{eq:K}
\end{equation}
This is the pivotal result of this article, a new type of FDR, connecting SVFC to SCN. 

%-----------------GENERAL PROPERTIES-----------------------------------------
\section{SVFC - general properties} \label{sec:svfc_gen_prop}
Let us discuss the general properties of SVFC. First thing is the limit of $r\to +\infty$, in which case the equation \eqref{eq:r2} reduces to $\gamma\dot r=\sqrt{2}\sigma\eta_{-}$, so the classical relation $\gamma=\beta \sigma^2/2$ holds. Since we expect that $U(r\to+\infty)\to0$, $g_{-}(r\to+\infty)\to1$ and the integral in the denominator of \eqref{eq:K} vanishes, we conclude that in this limit $K(r)\to\gamma$. 

Another issue is the limit of $r\to0$, in which case $K(r)\to0$. On the one hand, this is caused by $g_{-}(r\to 0)\to 0$, because of $h(r\to0)\to1$. On the other hand, for the particles interacting via the strongly repulsive core we expect $U(r\to0)\to+\infty$ and so the Boltzmann factor in \eqref{eq:K} disappears. $K(r)\simeq 0$ in the low-$r$ limit means that our over-damped theory leads to the necessarily frictionless hence inertial dynamics. This prediction has been recently confirmed by the analysis of the two-particle Mori-Zwanzig model \cite{bib:majka0}, which acts as the short-distance limit of the dynamics discussed here. Indeed, in the Mori-Zwanzig model the equation of motion for relative distance $r$ is also shown to be fully inertial and deterministic, with no friction nor stochastic force present \cite{bib:majka0}.% The same happens for short $r$ in \eqref{eq:r2}, where $K(r)$ diminishes and the noise is suppressed by $g_{-}(r\to0)\to0$.

Yet another aspect is the non-correlated case of $h(r)=0$ and $g_{\pm}(r)=1$. In this limit \eqref{eq:K} turns into:
\begin{equation}
K(r)=\frac{e^{-\beta U(r)}}{\frac{1}{\gamma}-\frac{2}{\beta\sigma^2}+\frac{2}{\beta \sigma^2}e^{-\beta U(r)}}=\gamma
\end{equation}
This shows that in the non-correlated limit our theory reduces to the ordinary Brownian diffusion, as expected.

Since SVFC $K(r)$ has been postulated in \eqref{eq:r2}, one might ask what are the equations of motion in absolute variables $x_i$. By the identity \eqref{eq:eta_def} we ca rewrite \eqref{eq:r2} and \eqref{eq:R2} in the form:
\begin{gather}
K(r)\dot r=2F(r)+\xi_2-\xi_1\\
2\gamma\dot R=\xi_2+\xi_1
\end{gather}
Summing and subtracting these equations we obtain the equations for $x_1$ and $x_2$:
\begin{gather}
\frac{1}{2}(\gamma+K(r))\dot x_1+\frac{1}{2}(\gamma-K(r))\dot x_2=-F(r)+\xi_1 \\
\frac{1}{2}(\gamma+K(r))\dot x_2+\frac{1}{2}(\gamma-K(r))\dot x_1=F(r)+\xi_2 
\end{gather}
where $r=x_2-x_1$ is kept for brevity. One can interpret these equations as an over-damped dynamics with the dissipative term $(\gamma+K(r))\dot x_i/2$ and an additional 'response force' $(\gamma-K(r))\dot x_i/2$. In the limit of huge $r$, where $K(r)\to\gamma$, these response forces vanish. However, for small $r$, even for $K(r)\to0$ the friction term is non-zero and the system remains dissipative.

Finally, from the structure of \eqref{eq:K} one concludes that $K(r)$ can become singular for a certain critical distance $r_c$, which satisfies:
\begin{equation}
\beta\int_{r_c}^{+\infty}dr'\frac{F(r')}{g_{-}(r')}e^{-\beta U(r')}=1
\end{equation}
Providing a general solution to this equation is difficult, but it might be satisfied e.g. for the purely repulsive potentials (i.e. $\forall_r F(r)\ge0$). In the vicinity of $r_c$ the equation \eqref{eq:r2} is dominated by the friction term, which becomes divergent, i.e. $K(r\to r_c^-)\to-\infty$, but $K(r\to r_c^+)\to+\infty$. This means that  for $r<r_c$ the term $K(r)\dot r$ turns into the source of active propulsion. This leads to a peculiar behavior, namely, we can rewrite \eqref{eq:r2} as:
\begin{equation}
\dot r = \frac{2F(r)}{K(r)}+\frac{\sqrt{2}\sigma g_{-}(r)}{K(r)}\eta_{-} \label{eq:r3}
\end{equation}
Linearizing this equation in the vicinity of $r_c$, we obtain:
\begin{equation}
\dot r\simeq -(r-r_c) \frac{K'(r_c)}{K^2(r_c)}\left( F(r_c)+\sqrt{2} \sigma g_{-}(r_c) \eta \right)
\end{equation}
where $K'(r_c)/K^2(r_c)=-\beta F(r_c)/(\gamma g_{-}^2(r_c))$. The analytical solution of this equation reads:
\begin{equation}
r(t)=r_c+(r_0-r_c) \exp\left( \frac{2\beta F^2(r_c)t}{\gamma g_{-}^2(r_c)}+\frac{\sqrt{2}\sigma\beta F(r_c)}{\gamma g_{-}(r_c)}W_t\right) 
\end{equation}
where $W_t$ is the Wiener process and $r_0$ is the initial position. The average value of $r(t)$ with respect to $W_t$, reads:
\begin{equation}
<r(t)>=r_c+(r_0-r_c)\exp\left( \frac{4\beta F^2(r_c)}{\gamma g_{-}^2(r_c)}t\right) \label{eq:delta_r}
\end{equation}
One can see that $<r(t)>$ bifurcates depending on $r_0$, i.e. for $ r_0>r_c$ the average distance can only grow in time, but for $r_0<r_c$ it systematically decreases. This means that for $r_0<r_c$ the two repulsive tracers can behave like an attractive system. These calculations are valid only locally, in the vicinity of $r_c$. However, from \eqref{eq:r3} one can see that for $F(r)>0$ and $K(r<r_c)\to0^-$ the $F(r)/K(r)$ term becomes an extremely amplified attractive force (though accompanied by the equally amplified noise term $g_{-}(r)\eta_{-}/K(r)$). We are not able to analyze this behavior beyond the linear approximation, but it might indicate the ergodicity breaking, i.e. the time evolution of the probability distribution for this system (formally given by the non-stationary FPE) might bifurcate. Namely, for the initial condition $r_0<r_c$ it might lead to the non-Boltzmann distribution in the long time-regime. A similar situation is encountered e.g. in the noisy logistic model \cite{bib:logistic} or in Ref. \cite{bib:denisov}, i.e. the same Langevin equation describes a few very different regimes of behavior, depending on parameters and initial conditions. In our case, we have used the stationary limit to determine the Langevin equation \eqref{eq:r2}, but now we use this equation to analyze the non-stationary limit, in which it must be also applicable. 

Nevertheless, the attraction effect predicted by the over-damped theory might not fully manifest in the physical systems, since for $K(r)\simeq 0$ the inertial behavior must be taken into account. The inclusion of inertial dynamics turns the attraction into the transient effect, as we show in the numerical calculations in the following section.

%-------------DLVO-EXAMPLE ANALYSIS------------------------------
\section{DLVO model: frictionless regime and transient attraction induced by SCN}\label{sec:DLVO}
Now, we apply our theory to the system of two charged spheres in the presence of counter-ions. According to the DLVO theory \cite{bib:lekkerkerker,bib:majka3}, the spheres interact via the Yukawa potential:
\begin{gather}
U_{Y}(r)=\frac{Q^2}{\epsilon (1+\kappa d/2)^2}\frac{e^{-\kappa(r-d)}}{r}\\
\kappa^2=\beta\frac{4\pi}{\epsilon}(n_1Q^2+n_2q^2)
\end{gather}
where: $Q$ - surface charge, $q$ counter-ion charge, $n_1$, $n_2$ - molar concentration of, respectively, spheres and  ions, $d$ - sphere diameter. In fact, $U_Y(r)$ is a valid effective potential \cite{bib:majka3}, so we divide the total sphere-sphere interaction $U(r)$ into the hard-core part $U_0(r)$  (i.e. $+\infty$ for $r\le d$, 0 otherwise) and effective part $U_{eff}(r)=U_Y(r)$. For calculations we choose the parameters from the typical colloidal range \cite{bib:crocker}: $Q=2000e$, $q=1e$, $d=600$ nm $n_1=2$ $\textrm{mm}^{-3}$, $T=300K$, $\gamma\simeq4.5\times 10^{-9}$kg/s (corresponding to the Stokes law in water) and the mass of spheres $m\simeq 4.5\times10^{-16}$kg. 

%--------------FIGURE 1------------
\begin{figure}
\includegraphics[width=0.9\columnwidth]{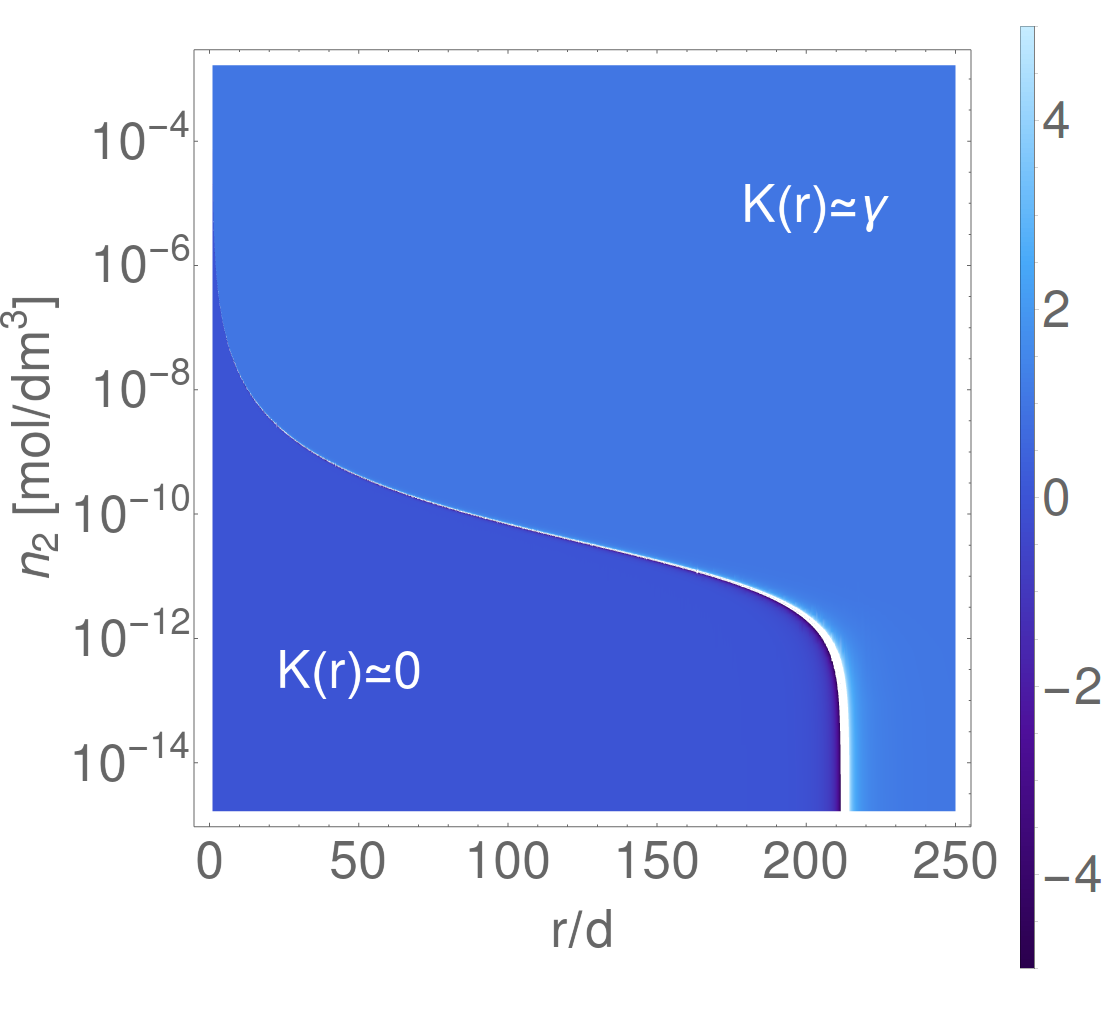}
\caption{The density plot of  $K(r)/\gamma$ for the DLVO-based model as a function of tracer-tracer distance $r$ and the counter-ion concentration $n_2$. The region of inertial motion ($K(r)\simeq 0$) can be clearly distinguished from the region of ordinary over-damped motion where $K(r)\simeq \gamma$. The two areas are separated by the narrow region of divergent behavior, $K(r)\to\pm\infty$ (see Fig. \ref{fig:K_r}). \label{fig:density_plot} }
\end{figure}
%------------FIGURE 2--------------
\begin{figure}
\includegraphics[width=0.95\columnwidth]{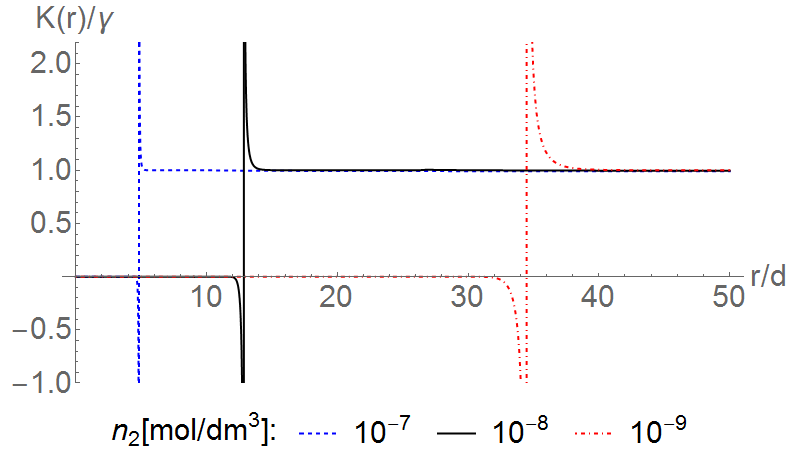}
\caption{The exemplary plots of $K(r)/\gamma$ in the DLVO-based model for the three values of $n_2$. $K(r)$ becomes divergent as it passes from $K(r)\simeq0$ to $K(r)\simeq \gamma$. \label{fig:K_r} }
\end{figure}

Knowing that $U_{eff}(r)=U_Y(r)$, we can numerically calculate the correlation function $h(r)$ from \eqref{eq:spat_cov} and \eqref{eq:spat_corr} to eventually predict $K(r)$. The numerical calculations confirm the general properties of SVFC discussed in the previous section. In the Fig. \ref{fig:density_plot} the density plot of $K(r)$ is presented as the function of $n_2$ and the distance $r$. For huge $r$ and $n_2$ the SVFC quickly saturates at $K(r)\simeq \gamma$. This is the region of the ordinary overdamped dynamics. Conversely, for low $n_2$ and $r$ the region of $K(r)\simeq 0$ occurs, which is governed by the inertial and frictionless dynamics. Since our model considers the spheres and counter-ions only, the region of $K(r)\simeq0$ can be interpreted as the counter-ion depleted zone. These two regions are separated by a narrow area in which $K(r)$ becomes singular, i.e. $K(r)\to\pm\infty$. This is illustrated in more detail in Fig \ref{fig:K_r}. We interpret the peak in $K(r)$ as the manifestation of a counter-ion rich layer. The concentration $n_2$ acts as the order parameter, because for $n_2\gtrsim 2\times10^{-5} \textrm{mol}/\textrm{dm}^3$ we have $r_c<d$. This means that the frictionless region is not accessible for higher $n_2$ and it emerges in a discontinues manner, marking the first-order phase transition. This effect is a consequence of a finite (i.e. non-point-like) tracer size. 

\begin{figure}
\includegraphics[width=0.95\columnwidth]{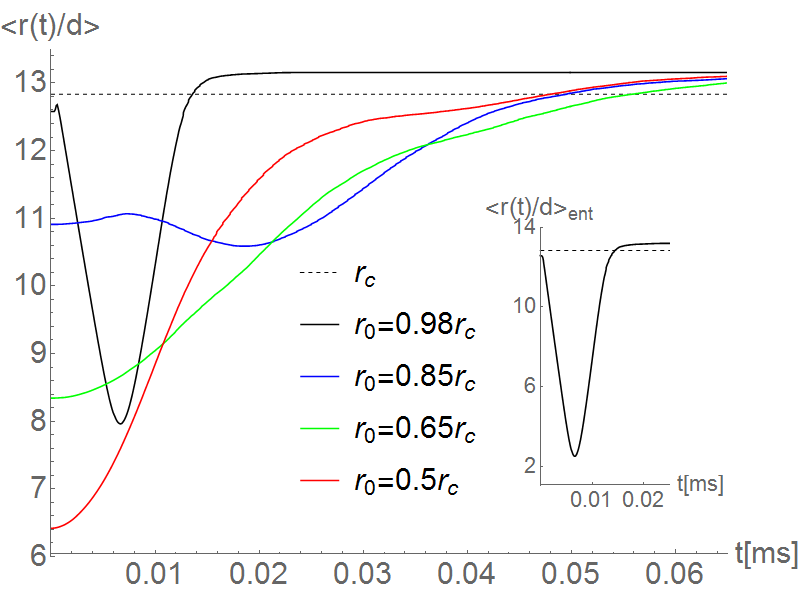}
\caption{The mean numerical solution $<r(t)>$ of equation \eqref{eq:full_dyn} for $n_2=10^{-8}$mol/$\textrm{dm}^3$ ($r_c=12.84d$, dashed line) in $T=300K$ and for several initial positions $r_0$. Despite the highly repulsive potential $U_Y(r)$, for $r_0\lesssim r_c$ the system exhibits systematic attraction in the initial phase of motion (especially: $r_0=0.98r_c$, black solid line), marking the transient attraction effect. Inset: for $r_0=0.98r_c$ the trajectories are either of the escape-type (54\% of cases) or the entrant-type (46\%). $<r(t)>_{ent}$ is calculated over the entrant trajectories and indicates the depth of penetration in the attraction effect. \label{fig:trajectory}}
\end{figure}

Finally, we examine whether the attractive behavior predicted by \eqref{eq:delta_r} occurs. In the light of the discussion given at the end of Sec. \ref{sec:svfc_gen_prop}, we have decided to perform the numerical simulations of the full dynamics: 
\begin{equation}
m\ddot r +K(r) \dot r = 2 F(r)+\sqrt{2}\sigma g(r)\eta \label{eq:full_dyn}
\end{equation}
where $K(r)$ is still calculated from the over-damped model \eqref{eq:K}. We choose $n_2=10^{-8}$ mol/$\textrm{dm}^3$ ($r_c\simeq12.84d$). In general, $m/\gamma=10^{-7}s$, but we have also checked that $m/|K(r)|<0.1s$ for $r>11.22d$ and it rapidly decreases to 0 at $r_c$. This justifies the over-damped dynamics in the vicinity of $r_c$. In the Fig. \ref{fig:trajectory} we show $<r(t)>$ calculated from \eqref{eq:full_dyn} for several sets of initial conditions. The average is taken over 500 trajectories with initial velocity $v_0=0$ and initial position $r_0$ ranging from $0.5r_c$ to $0.98r_c$. For trajectories starting in the close vicinity of $r_c$ (e.g. $r_0=0.98r_c$), where $K(r)<0$, the particles experience strong propulsion, which allows them to either leave the anomalous region immediately (escape trajectories) or to penetrate this region (entrant trajectories). In the latter case, after the particles experience the propulsion, $r(t)$ decreases in a frictionless manner until the electrostatic repulsion prevails. It is in this phase that without the weak inertial term in \eqref{eq:full_dyn} $r(t)$ would fall below 0, which would be unphysical. The particles are then expelled into the high friction region ($r>r_c$) and in the vicinity of $r_c$ they are propelled once again by $K(r)<0$, but this time this propulsion facilitates leaving. This entire phase lasts approximately 10-20[$\mu$s] and this is what we call the transient attraction effect. 

When the average is taken solely over the entrant trajectories (46\% of cases for $r_0=0.98r_c$, the inset plot in Fig. \ref{fig:trajectory}), it is revealed that the minimal distance $r_{min}$ can be as small as $2.5d$ on average, but the trajectories with $r_{min}\simeq d$ are also observed. On average, the transient attraction effects weakens with decreasing $r_0$ and it does not manifest in $<r(t)>$ for $r_0<0.65r_c$. However, for such $r_0$ many trajectories exhibit multiple attraction-repulsion cycles, i.e. when they return to the propelling region they are pushed back instead of leaving. Nevertheless, even these trajectories eventually escape from the anomalous region. In general, we have found no indication that the particles can stay in the $r<r_c$ region permanently or for a significantly long time. Basing on these observations we describe the attraction effect as transient. We should also comment that for $n_2=10^{-8}$mol/$\textrm{dm}^3$ the energy barrier reads $U_Y(r_c)\simeq 6kT$ hence it is extremely unlikely that the expelled particles ($r>r_c$) reenter the anomalous region spontaneously.

%--------------------FIGURE 3--------------------

%-------------------------CAVEATS-------------------
\section{Caveats and comments}\label{sec:caveats}
The results we present raise several methodological questions. First of all, there is a general problem of whether the SCN-driven Langevin dynamics can be applied to the non-equilibrium regime. A single ordinary Langevin equation describes both the stationary and transient states. It is the initial condition that decides to which regime its solution (a trajectory) belongs. This is also true in our case. Yet, since the tracer-bath coupling is spatially correlated, the dynamics including SCN seems more physically adequate than one that neglects it. However, our Langevin equation explicitly depends on the spatial correlation function $h(r)$, which has been derived for equilibrium conditions. Thus, our dynamics can work only in the 'close to equilibrium' limit, where the non-equilibrium correlations are well approximated by the equilibrium one. Unfortunately, up to our knowledge, no efficient alternative currently exists, since the direct microscopic approach such as e.g. the Mori-Zwanzig model leads to a very limited analytical insight \cite{bib:majka0}.

Another issue is that in this paper we postulate SVFC as a function of $r$ only. In fact, this choice is arbitrary as one could consider models with SVFC dependent on the velocities as well. However, such theory would be substantially more complicated, so the current model must be seen as an intermediate step towards a more versatile solution. This is also important in the context our simulations, in which we eventually combine the over-damped theory for $K(r)$ with the inertial dynamics. While this seems physically justified, these two aspects are not guaranteed to be compatible in the current approach. We have also no quantitative criterion to assess the validity of results. Therefore, the transient attraction effect must be treated with a huge dose of caution. These problems could be resolved by the inertial theory.

Finally, let us comment on the relation to experiment. Except for Ref. \cite{bib:like3}, no experimental work known to authors is oriented solely on the transient aspects of the screened-charge attraction. While the escape/entrant behavior which we simulate resembles the transient observations from Ref. \cite{bib:like3}, this experiment deals with significantly different length- and time-scales. It is also affected by a number of factors (e.g. the sphere-surface interaction, tapping procedure, hydrodynamics effects etc.) which are beyond the scope of our theory and make the direct comparison impossible. We should also emphasize that the transient attraction effect cannot explain the like-charge attraction in the equilibrated systems, which must involve the effective interactions essentially different from the purely repulsive DLVO potential. However, the transient attraction effect might accompany the like-charge attraction in the equilibration phase. Therefore our results might prompt additional interests in the dynamics of such systems.

\section{Summary}
We have shown that the spatially correlated noise is an ubiquitous component of the molecular world and we have proposed the thermodynamically consistent Langevin dynamics with SCN and its appropriate FDR. Our results indicate that the simultaneous diffusion of many particles driven by a common thermal bath can include collective effects. The application to the DLVO-based model shows that SCN can induce qualitatively new dynamics, such as the emergence of the frictionless regime and the transient attraction effect. These effects might facilitate the self-assembly of particles which can bind chemically, but repel in the electrostatic manner and provide an auxiliary transportation mechanism. This is especially interesting in the context of engineered soft-matter systems and the problem of protein complexation. However, the applicability of our theory is currently difficult to assess and a further development in this field is necessary. Especially, the formal extension to the inertial regime is of utmost interest. 

\begin{acknowledgments}
The authors gratefully acknowledges the National Science Center, Poland for the grant support (2014/13/B/ST2/02014).
\end{acknowledgments}

\end{document}